\documentclass[aps,preprint,showpacs,amssymb,amsmath,nofootinbib]{revtex4}

\usepackage{graphicx}

\newcommand{\mcs}{\mathcal{S}}
\newcommand{\psibar}{\overline{\psi}}
\newcommand{\Psibar}{\overline{\Psi}}
\newcommand{\half}{\frac{1}{2}}

\DeclareMathOperator{\sech}{sech}

\begin{document}


\title{Fermions, scalars and Randall-Sundrum gravity on domain-wall branes}
\author{Rhys Davies}
\email{daviesr@physics.unimelb.edu.au}
\author{Damien P. George}
\email{d.george@physics.unimelb.edu.au}
\affiliation{School of Physics, Research Centre for High Energy
Physics,\\The University of Melbourne, Victoria 3010, Australia}
\date{\today}

\begin{abstract}

We analyse the general features of localisation of fermions and
scalars in smoothed field-theoretical versions of the type 2 Randall-
Sundrum braneworld model.  A scalar field domain-wall forms the brane,
inducing warped gravity, and we study the mass spectra of the matter
fields in the dimensionally reduced theory.  We demonstrate
explicitly that both scalar and fermion fields exhibit a continuum
of properly normalisable modes starting at zero mass.  If discrete bound
modes are present in the gravity-free case, these become resonances
in the continuum, while off-resonant modes are highly suppressed on the
brane.  We describe briefly how another scalar field can be used to break a
symmetry on the domain-wall while leaving it unbroken far from the wall,
as has already been done in the flat space case.
Finally we present numerical calculations for a toy model which demonstrates
the decoupling of continuum modes at low energies, so the theory becomes
four dimensional.

\end{abstract}


\maketitle


\section{Introduction}

Over the last decade there has been a lot of research into the
possibility that our observable $3+1$-dimensional universe resides
on a defect in a higher-dimensional bulk spacetime.  If this is
the case, we should be able to construct a model which explains
why physics at low energies is insensitive to the extra dimensions.
It has been known for some time that bulk fermions coupled to a
scalar field domain-wall have a spectrum containing a zero
mode which is localised to the wall~\cite{Rub&Shap}.  In the case of a
$4+1$-dimensional bulk, this zero mode fermion is chiral and
separated by a mass gap from the massive modes, making it
well suited to model building efforts.  Scalar fields can be localised
on the wall in a similar fashion, and in particular, the canonical Mexican hat
potential can be generated for the lowest lying mode~\cite{George}.

The gravitational part of the puzzle found a solution in the
type 2 Randall-Sundrum model (RS2)~\cite{RS2}, where the metric is
warped to ensure $3+1$-dimensional gravity is
reproduced on an infinitely thin defect.  Extensions to smoothed
out defects induced by scalar fields have since been
constructed (see eg.~\cite{DeWolfe1999, Gremm1999, Davidson2000, Volkasclash, KT, Slatyer2007}).
When combining this warped gravity set up with the fermion
and scalar localisation, it is important to ensure that gravity
doesn't destroy the desirable features of the matter localisation.

For the case of scalar fields, it was first noted in~\cite{Bajc1999}
that with gravity included, the scalar continuum modes
can begin at zero energy.  In~\cite{Dubovsky2000} it was shown that
this can be true for fermions also, and that introducing a $4+1$-%
dimensional mass term can produce massive, localised, meta-stable
states.  Due to their coupling to the low lying continuum, these
states can tunnel into the bulk and have a finite lifetime.
Ref.~\cite{Ring} considered a specific model and determined the
full mass spectrum of the meta-stable, or quasi-localised, modes
and demonstrated that their lifetime could be made longer the age
of the universe.

In this paper we present a general analytic argument which
demonstrates explicitly that in the presence of RS2-like gravity, the
continuum matter modes always begin at zero energy.  It is
also demonstrated that the discrete modes present in the
gravity-free theory show up in the warped case as resonances of
the domain-wall trapping potential, which represent the quasi-localised
states.
It can be argued that in the presence of interactions, the low-energy non-resonant
continuum modes will couple only very weakly to modes localised on the
brane.  Thus $3+1$-dimensional physics can still be reproduced at low energies.
These considerations are important for realistic model-building~\cite{Davies2007}.

The paper is structured as follows.  In Section~\ref{sec:ferm-no-grav}
an overview of localised fermions in the gravity-free case is given, 
followed by an analysis with gravity included in
Section~\ref{sec:ferm-grav}.  This later section demonstrates that
the fermion mass spectrum has a continuum beginning at zero, and
that these massive modes can be properly normalised in the presence
of the warped metric.  In Section~\ref{sec:scalar-grav} we repeat
the gravitational analysis for a coupled scalar field and
show that while the continuum modes begin at zero mass,
it is still possible to obtain discrete modes with tachyonic mass,
allowing one to realise the Higgs mechanism.
In Section~\ref{sec:toy-model} we consider a specific toy model to
demonstrate more clearly the effect of gravity,
and provide numerical support to the claim that the gravity induced
continuum modes are only weakly coupled to a brane-localised zero
mode.  Section~\ref{sec:concl} concludes.


\section{Fermions in the gravity-free case}
\label{sec:ferm-no-grav}

We begin with a brief review of the situation without gravity.
Consider a 5D model containing a number of scalar fields $\Phi_j$.
These will form our classical background.  The action is,
\begin{equation} \label{flataction}
\mcs_{bg} = \int \! d^4 x \! \int \! dy \,\\
	\left \{ \half \partial^A \Phi_j \partial_A \Phi_j - V(\Phi) \right \}.
\end{equation}
Here $A=(0,1,2,3,5)$ is a 5D Lorentz index and repeated indices
are summed.  The only additional assumptions we make are that $V$ has
a $\mathbb{Z}_2$ symmetry $\Phi_j \to -\Phi_j \; \forall \; j$,
such that the global minimum of $V$ is at least doubly degenerate,
and attained for (say) $\Phi_j = \pm \Phi^{\min}_j$.  We also require this
$\mathbb{Z}_2$ to be independent of any continuous symmetries of
the theory.

We suppose we have a solution $\phi_j(y)$ depending only on the
coordinate $y$, satisfying the boundary conditions $\phi_j \to \pm
\Phi^{\min}_j$ as $y \to \pm \infty$.  Such a solution is
topologically stable, and can be used as a classical background
for a quantum field theory~\cite{Raj}.

Now introduce a fermion field $\Psi$ into the model, and Yukawa-couple
it to the domain-wall.  Its action will be,
\begin{equation}
\mcs_\Psi = \int \! d^4 x \! \int \! dy
    \, \left \{ i \Psibar \Gamma^A \partial_A \Psi
	- g_j \Phi_j \Psibar \Psi \right \},
\end{equation}
where the $g_j$ are Yukawa coupling constants and
$\Gamma^\mu = \gamma^\mu, \Gamma^5 = - i \gamma^5$ with
$\gamma^{\mu, 5}$ the usual 4D Dirac matrices and chirality
operator, respectively.  The action of the $\mathbb{Z}_2$ symmetry
is extended to include $y \to -y$ and $\Psi \to \Gamma^5 \Psi$.
For simplicity we have also imposed a global $U(1)$ symmetry $\Psi\to
e^{i\theta}\Psi$, to forbid a term $g'_j\Phi\Psibar\Psi^c + h.c.$.
Therefore, in the classical background discussed above, the Dirac
equation will be,
\begin{equation}
\left[ i \Gamma^A \partial_A - g_j \phi_j(y) \right] \Psi(x^\mu, y) = 0.
\label{eq:dirac}
\end{equation}
To solve this equation, we separate variables by expanding $\Psi$
in a generalised Fourier series,
\begin{equation}
\Psi(x^\mu, y) = \sum_{n} \left[ f^n_L(y) \psi^n_L (x^\mu) + f^n_R (y) \psi^n_R (x^\mu) \right],
\end{equation}
where the $\psi_{L, R}$ are left- and right-handed 4D spinors, and
are treated independently due to the association of $\Gamma^5$ with
$y$.  The sum over $n$ generally includes an integral over
continuum parts.  We choose our basis functions such that the
$\psi_{L, R}$ satisfy the 4D Dirac equation,
\begin{align*}
i \gamma^\mu \partial_\mu \psi^n_L = m_n \psi^n_R \quad \text{and}
	\quad i \gamma^\mu \partial_\mu \psi^n_R = m_n \psi^n_L.
\end{align*}
We can then solve Eq.~\eqref{eq:dirac} for $f^n_{L,R}$.  In the
$m_0 = 0$ case, we get the decoupled equations (hereafter, primes
denote differentiation with respect to $y$),
\begin{equation}
\begin{aligned}
{f_L^0}' + g_j \phi_j f_L^0 &= 0, \\
{f_R^0}' - g_j \phi_j f_R^0 &= 0.
\end{aligned}
\end{equation}
The solutions are,
\begin{equation}
f^0_{L, R}(y) \propto \exp \left ( \mp \int^y \! d\tilde{y} \, g_j \phi_j(\tilde{y}) \right ).
\end{equation}
The boundary conditions satisfied by the $\phi_j$ then give
generically that if $g_j \Phi^{\min}_j > 0$ the left-handed component
of $\Psi$ is localised near $y = y_0$, where $g_j \phi_j(y_0) = 0$
(the existence of such a point follows from the boundary conditions),
while the right-handed component is non-normalisable, and thus
unphysical.  This is the well-known result that a domain-wall localises
chiral fermions~\cite{Rub&Shap}. More interesting is the case $m_n > 0$.
We then get the following equation for $f_{L,R}$,
\begin{equation}
- {f_{L,R}^n}'' + W_\mp f_{L,R}^n = m_n^2 f_{L,R}^n, \label{eq:flatSchro}
\end{equation}
where $W_\pm = (g_j \phi_j)^2 \pm g_j \phi_j'$.  Eq.~\eqref{eq:flatSchro}
is just a Schr\"odinger equation with eigenvalue $m_n^2$.  The potential
$W$ is a finite well, and $W \to (g_j \Phi^{\min}_j)^2 > 0$ as
$y \to \pm \infty$.  Therefore the normalisable solutions will consist
of a number of discrete bound states, as well as a continuum starting
at some non-zero energy%
\footnote{The continuum modes will only be delta-function normalisable.}.
Thus the 4D fermion spectrum contains a
massless left-handed particle, a finite number of massive Dirac particles
with discrete masses, and a continuum of massive Dirac particles beginning
at $m_\text{cont} = g_j \Phi^{\min}_j$.  Explicit solutions to one specific
model can be found in~\cite{George}.


\section{Including gravity}
\label{sec:ferm-grav}

To include gravity in these models, we simply add the Einstein-Hilbert
term to the action, and minimally couple other fields to gravity as
usual, to obtain the background action
\begin{equation}
\mcs_{bg} = \int \! d^4 x \! \int \! dy \, \sqrt{G} \Big \{ -2 M^3 R - \Lambda
 + \half G^{M N} \partial_M \Phi_j \partial_N \Phi_j - V(\Phi) \Big \},
\label{eq:gravKink}
\end{equation}
where $G_{MN}$ is the 5D metric, $G$ its determinant, $M$ the 5D Planck
mass, $R$ the 5D Ricci scalar, and $\Lambda$ the bulk cosmological constant.

Now we must solve the coupled Einstein and Klein-Gordon equations associated with~\eqref{eq:gravKink}.
We again assume that the solution $\phi_j(y)$ depends
only on $y$, and satisfies the same boundary conditions as earlier.
The resulting equations have been solved by various
authors~\cite{DeWolfe1999, Gremm1999, Davidson2000, Volkasclash, KT, Slatyer2007}
and here we merely point out the general
features.  The solution for the metric is given by,
\begin{equation} \label{eq:metric}
ds^2 = e^{-2 \sigma(y)} \eta_{\mu \nu} d x^\mu d x^\nu - d y^2
\end{equation}
where $\sigma$ is a smooth even function of $y$, and $\sigma \to \mu |y|$ as
$|y| \to \infty$, where $\mu$ is some mass scale.  In the RS2 model, the
domain-wall is instead a delta-function brane, and $\sigma(y) = \mu |y|$ everywhere.
The effective 4D graviton spectrum, corresponding to fluctuations around
the metric in Eq.~\eqref{eq:metric}, was shown to consist of a single massless
mode, followed by a continuum of massive modes starting arbitrarily close
to zero mass~\cite{RS2}.
Contrary to na\"ive expectations, this does not contradict the assertion
that the low-energy theory is 4 dimensional.  In fact, the integrated
effect of the continuum modes at the position of the brane is
negligible at low energies, due to the suppression of their wavefunctions
near the brane.  Cs\'aki et al. have shown that the same result holds
in the present context of smooth RS-like spacetimes \cite{Csaki}.  We will
demonstrate in this and the next section that similar statements are true
for fermions and scalar fields coupled to the above background.

With gravity included, the formulation of the Dirac Lagrangian requires
the introduction of the vielbein $V_A^{N}$ (here $A$ is an `internal'
Lorentz index) and the spin connection $\omega_N$,
given for the metric in Eq.~\eqref{eq:metric} by,
\begin{equation}
\begin{aligned}
V_A^{\mu} & = \delta^{\mu}_A e^{\sigma} \quad &\omega_{\mu}
	& = \frac{i}{2} \sigma' e^{-\sigma} \gamma_{\mu} \gamma^5 \\
V_A^{5} & = \delta^5_A   &\omega_5 & = 0.
\end{aligned}
\end{equation}
These yield the spin-covariant derivative $D_N = \partial_N + \omega_N$,
and curved space gamma matrices $\Gamma^N = V^N_A \Gamma^A$, so that
the fermion action is,
\begin{equation} \label{eq:action}
\mcs_\Psi = \int \! d^4 x \! \int \! d y \, \sqrt{G}
    \big \{ i \Psibar \Gamma^N D_N \Psi - g_j \Phi_j \Psibar \Psi \big \}.
\end{equation}
The resulting Dirac equation is,
\begin{equation}
\left[ \gamma^5 \partial_y + i e^{\sigma} \gamma^\mu \partial_\mu
    - 2\sigma' \gamma^5 - g_j \phi_j(y) \right] \Psi(x^\mu,y) = 0.
\end{equation}

We again decompose $\Psi$ into 4D chiral components, and obtain
equations for the extra-dimensional profiles $f_{L,R}$.  As long
as the value $g_j \Phi^{\min}_j$ is large enough, the conclusions
are unchanged in the massless case (see eg.~\cite{Slatyer2007, Bajc1999}).
For $m_n > 0$ we get the following equations,
\begin{equation} \label{eq:quasiSchro}
\begin{aligned}
& -{f_{L,R}^n}'' + 5\sigma' {f_{L,R}^n}' 
    + \left[ 2 \sigma'' - 6\sigma'^2 + \tilde{W}_\mp \right] f_{L,R}^n
        = m_n^2 e^{2 \sigma} f_{L,R}^n,
\end{aligned}
\end{equation}
where $\tilde{W}_\pm = (g_j \phi_j)^2 \pm g_j \phi_j' \mp g_j \phi_j \sigma'$.
The inclusion of gravity has meant that Eq.~\eqref{eq:quasiSchro} is
no longer simply a Schr\"odinger equation, so we can't directly
apply our knowledge of 1D quantum mechanics.  It is however possible to
transform Eq.~\eqref{eq:quasiSchro} into a Schr\"odinger equation.
Specifically, we let $f_{L,R}^n = e^{2 \sigma} \tilde{f}_{L,R}^n$,
and change coordinates to $z(y)$ such that
$\frac{dz}{dy} = e^{\sigma}$ (this is in fact a change to `conformal
coordinates', in which $ds^2 = e^{-2 \sigma(y(z))}(\eta_{\mu \nu}
dx^\mu dx^\nu - dz^2)$).  Eq.~\eqref{eq:quasiSchro} becomes,
\begin{equation} \label{eq:curvedSchro}
\left[-\frac{d^2}{d z^2}
    + e^{-2 \sigma} \tilde{W}_\mp \right] \tilde{f}_{L,R}^n
    = m_n^2 \tilde{f}_{L,R}^n.
\end{equation}
We thus identify the effective potential,
\begin{equation} \label{eq:effpot}
\tilde{W}_\pm^{\text{eff}} = e^{-2 \sigma} \tilde{W}_\pm.
\end{equation}
As $|y| \to \infty$, $\sigma \sim \mu |y|$, which in terms of $z$
becomes $e^{-2 \sigma} \sim 1/(\mu z)^2$ as $|z| \to \infty$.
As $|z| \to \infty$, we have $\tilde{W}_\pm \to \text{constant}$, and
therefore the effective potential decays towards zero at
large distances from the brane (see Fig.~\ref{fig:pot} for some specific
cases).  Indeed, it is an example of a `volcano potential', familiar from
analysis of the graviton sector~\cite{Gremm1999}.
Particles subjected to $\tilde{W}_\pm^{\text{eff}}$ are essentially free asymptotically, so
there is a continuum of delta-function normalisable solutions for all $m_n^2 > 0$.
We will now show that normalisability of $\tilde{f}_{L, R}^n$ implies appropriate
normalisability of $f_{L, R}^n$, and conclude that our dimensionally reduced theory
contains a continuum of fermions starting at zero mass.

The $\tilde{f}_L^n$ satisfy an ordinary Schr\"odinger equation with a
continuum of eigenvalues, and therefore are delta-function orthonormalisable,
\begin{equation}
\int_{-\infty}^{\infty} \! d z \, \tilde{f}_L^{n*} \tilde{f}_L^{n'}
    = \delta(n - n').
\end{equation}
On the other hand, the normalisation condition for the $f_{L, R}^n$ can be
derived by demanding that integrating the action in Eq.~\eqref{eq:action}
over $y$ leads to a properly normalised 4D kinetic term, viz.
\begin{align}
\int \! d ^4 x \! \int \! d y \sqrt{G} \left \{ i \Psibar \Gamma^A V_A^{\mu}
    \partial_\mu \Psi \right \}
    = \int \! d^4 x \! \int \! d n  \, i \, \psibar^n \! \gamma^\mu
    \partial_\mu \psi^n.
\end{align}
Substituting in the expressions for the vielbein and metric, this condition becomes,
\begin{equation}
\int_{-\infty}^\infty \! d y \, e^{-3 \sigma} f_L^{n*} f_L^{n'} = \delta(n - n'),
\end{equation}
and similarly for the right-handed components (overlap integrals between
right- and left-handed components do not enter).  However, if we write
$f_L^n$ in terms of $\tilde{f}_L^n$ and use $dy = e^{-\sigma} dz$, we see that,
\begin{equation}
\int_{-\infty}^\infty \! d y \, e^{-3 \sigma} f_L^{n*} f_L^{n'}
= \int_{-\infty}^{\infty} \! d z \, \tilde{f}_L^{n*} \tilde{f}_L^{n'}.
\end{equation}
So the normalisation integral for the $f_L^n$ is equivalent to the
normalisation integral for the $\tilde{f}_L^n$.  Therefore there
is a continuum of normalisable fermion modes in the theory, starting
at zero 4D mass.  Despite this, the zero modes still form an effective
4D theory at low energies.  We can understand this as follows.

Flat space corresponds to $\sigma \equiv 0$, and we have seen that
in this case the low-energy spectrum consists of a finite
number of particles with discrete masses.  The reason for this is
that the effective potential of the analogue Schr\"odinger system,
given in Eq.~\eqref{eq:flatSchro}, asymptotes to a non-zero value;
the discrete spectrum corresponds to modes bound in the potential
well near $y = 0$.

Suppose now that $\sigma$ is non-zero, but grows only very slowly
with $|y|$.  In this case, the effective potential in
Eq.~\eqref{eq:curvedSchro} approximates that of
Eq.~\eqref{eq:flatSchro} near the brane, then decays towards zero
as $|y| \to \infty$.  We thus have a localised non-zero potential in
the form of a narrow well flanked by wide barriers.  The low-energy
continuum eigenfunctions of such a system will generically have very
small amplitudes at the position of the well, due to the potential
barrier which they must tunnel through.

So although arbitrarily light fermions will exist in the theory, their
wavefunctions will be strongly suppressed at the position of the brane,
where the zero modes reside.  They are effectively `localised at
infinity'.  This leads to a very small probability of these low-energy
continuum modes interacting with the zero modes.

There is one more generic feature which we expect to occur.  Certain
discrete energies will resonate with the potential, and the
corresponding states will thus have a much larger probability of being
found on the brane.  These are the remnants of the discrete bound
states in the flat space case, and become coincident with them in the
zero-gravity limit.  What happens if one of these resonant modes is
produced in a high-energy process on the brane?  Any particle
produced on the brane will have a wavefunction truly localised to the
brane, and thus cannot correspond exactly to a single mode $\psi^n$,
which has a wavefunction oscillatory as $z\to\infty$.  Instead it will
be a wavepacket made from the continuum modes, with a Fourier spectrum
peaked around one of the resonances.  Therefore it is not a true
energy/mass eigenstate, and as the various components become out of
phase, the wavefunction will leak off the brane.  The particle then
has some probability of escaping the brane, which justifies the
moniker ``quasi-stable" or ``quasi-localised" for the resonant modes.
It is these resonant quasi-localised states that are investigated
in~\cite{Ring}.

Quantitative calculations confirming the above conclusions will be
given for a particular toy model in Section~\ref{sec:toy-model}.


\section{Localised scalar fields with gravity}
\label{sec:scalar-grav}

As well as fermions, it is desirable for model building purposes
to be able to localise scalar fields to the wall.  In flat space,
the results are similar to those for fermions, except that the
mass-squared of the lightest mode depends on parameters in the 5D
theory~\cite{George} (whereas in the fermion case, it is always zero).
It can even be arranged to be negative, so as to realise the Higgs
mechanism in the low-energy theory. We will now examine the effects
of gravity on these results.

We consider a scalar field $\Xi$ described by the action,
\begin{equation} \label{eq:scalar-act}
\mcs_\Xi = \int \! d^4 x \! \int \! d y \, \sqrt{G} \left \{ G^{MN} (\partial_M \Xi)^\dagger \partial_N \Xi
- H(\Phi, \Xi) \right \},
\end{equation}
where $H$ specifies the coupling of $\Xi$ to itself and to the
domain-wall.  The linearised equation of motion for $\Xi$ is given by,
\begin{equation}
\partial_M \left ( \sqrt{G} \, G^{MN} \partial_N \Xi \right ) + \sqrt{G} \, U (\phi_j) \Xi = 0,
\end{equation}
where $U$ is independent of $\Xi$, and defined by
$\frac{\partial H}{\partial \Xi^\dagger} = U \Xi + \mathcal{O}(\Xi^2)$.
We solve this exactly the same way as in the fermion case:
\begin{equation} \label{eq:scalarmodes}
\mbox{Let} \quad \Xi(x^\mu, y) = \sum_n h^n(y) \xi^n(x^\mu),
\end{equation}
where each $\xi^n$ satisfies a 4D Klein-Gordon equation,
\begin{equation}
\partial^\mu \partial_\mu \xi^n + m_n^2 \xi^n = 0.
\end{equation}
The analogue of Eq.~\eqref{eq:quasiSchro} is then,
\begin{equation}
-{h^n}'' + 4 \sigma' {h^n}' + U h^n = e^{2 \sigma} m_n^2 h^n
\end{equation}
We can convert this to a Schr\"odinger equation by again going to
the conformal coordinate $z$, as well as making the substitution
$h^n = e^{\frac{3}{2} \sigma} \tilde{h}^n$.  This yields,
\begin{equation} \label{eq:scalar-de}
-\frac{d^2 \tilde{h}^n}{d z^2} + \left [ -\frac{3}{2} \frac{d^2 \sigma}{d z^2}
	+ \frac{9}{4} \left ( \frac{d \sigma}{d z} \right )^2 \! + e^{-2 \sigma} U \right ] \tilde{h}^n
		= m_n^2 \tilde{h}^n.
\end{equation}
As $|z| \to \infty$, $\sigma \sim \log|z|$, and $U \to$ {\it constant},
so we can see immediately
that, as in the fermion case, the effective potential
decays towards zero far from the brane.  Therefore the
low-energy scalar spectrum also contains a continuum of modes of
arbitrarily small mass, which are properly normalisable, as can be
shown by a calculation analogous to that described above for the
fermions.

If $U \equiv 0$, the above equation is in fact identical to that
satisfied by 4D gravitons in the background of Eq.~\eqref{eq:metric}
(see eg. Ref~\cite{Csaki}).  In this case then, we know that
there is a single zero mode, followed by a continuum of modes starting
arbitrarily close to $m_n^2=0$.  The low lying continuum modes are
strongly suppressed on the brane; for example, their contribution to a
static potential generated by $\Xi$ exchange between two sources on
the brane separated by $r$, is suppressed by $1/(\mu r)^2$ relative
to the contribution of the zero mode.

For non-zero $U$, the spectrum is modified from the graviton case,
the significant difference being the possible introduction of
resonant modes (in the absence of fine-tuning of parameters, there
will no longer be a zero mode).  As in the fermion case, these
resonant modes correspond to the discrete bound modes in the
corresponding gravity-free theory and we expect the first of these
modes to occur for $m_n \sim \mu$.
Unlike the fermion case,
if appropriate coupling to the domain-wall is included, such that
$U$ makes some negative contribution to the effective potential,
then there may be bound state solutions with $m_n^2 < 0$, as in the
gravity-free case~\cite{George}.  This signals an instability in the
system, and implies that $\Xi$ is non-zero in the stable background
configuration.  In this case we would have to instead solve the
coupled Einstein and Klein-Gordon equations including the $\Phi_j$
fields {\it and} $\Xi$.
This setup can be used for interesting model building, in which a
symmetry is broken on the brane but restored in
the bulk.  This idea has been used in the flat space case in
Ref.~\cite{DvaliShifman}.  We sketch the reasoning following
Ref.~\cite{Eddie}.  Take the scalar potential
\begin{equation}
H(\Phi,\Xi) = (g'\Phi^2 - u^2)\Xi^\dagger\Xi + \tau(\Xi^\dagger\Xi)^2
\end{equation}
where we have specialised to a single background field $\Phi$, and
we assume $g'\Phi^{\min} - u^2>0$ such that $(\Phi,\Xi) = (\pm\Phi^{\min},
0)$ are still the global minima of the potential, and we must have
$\Xi\to0$ as $|y|\to\infty$.  If $\Phi$ forms a domain wall, then
$\Phi \sim 0$ inside the wall, so that the leading term of $H(\Phi,\Xi)$
is $\sim -u^2\Xi^\dagger\Xi$, suggesting that
the $\Xi=0$ solution is unstable there.  This will show up as a
negative eigenvalue $m_n^2 < 0$ in equation~\eqref{eq:scalar-de},
and solving for a consistent set of background solutions will
yield a background $\Xi$ that is peaked on the brane and tending
to zero in the bulk.  Putting $\Xi$ in a non-trivial representation
of some gauge group will induce spontaneous breaking of that group
on the brane, a mechanism which can be used, for example, to
realise the standard model Higgs mechanism on the brane.

In the stable case then, asymptotically $U$ will approach some
constant positive value $U_0$.  As $|z| \to \infty$, we can
approximate the effective potential $V_\text{eff}$ as follows:
\begin{equation}
V_\text{eff} \sim \frac{1}{z^2} \left ( \frac{15}{4} +
    \frac{U_0}{\mu^2} \right ).
\end{equation}
Again we can appeal to the results of Ref.~\cite{Csaki}, where
it is shown that for a potential that behaves asymptotically
as $\alpha (\alpha + 1)/z^2$, the amplitudes of modes with small $m_n$
are suppressed by $(m_n/\mu)^{\alpha - 1}$.  Therefore the coupling
to the domain-wall actually reduces the effect of the continuum modes
on low energy physics.


\section{Toy model calculation}
\label{sec:toy-model}

The above conclusions can be illustrated concretely by finding numerical
results for a specific case.  We will make use of a background
solution found in Ref.~\cite{Kobayashi2001} in which a single real
scalar field forms the domain-wall.  The stability of this solution
is demonstrated in Ref.~\cite{Kobayashi2001}. The solutions for the warp factor
and scalar field are\footnote{The functional form of this solution
appears different to that given in Ref.~\cite{Kobayashi2001}, but it
is in fact equivalent.},
\begin{equation} \label{eq:solution}
\begin{aligned}
\sigma(y) &= a \log(\cosh(ky))\\
\phi(y) &= D \arctan(\sinh(ky)),
\end{aligned}
\end{equation}
where $a = D^2/12 M^3$ is proportional to the 5D Newton's constant,
and the solution corresponds to a 5D cosmological constant given by
$\Lambda = -D^4 k^2/ 6 M^3$.  The scalar field potential which admits
this solution is,
\begin{equation}
V(\Phi) = 6 a k^2 M^3 (1 + 4a) \sin^2\left ( \frac{\phi}{D} -
    \frac{\pi}{2} \right ).
\end{equation}
We'd like to study a fermion field $\Psi$ in the above background to
illustrate the existence and suppression of the low-lying continuum
modes.  We again impose a global $U(1)$ symmetry $\Psi\to e^{i\theta}
\Psi$ so that $\Psi$ only couples to the background via a term
$g\Phi\Psibar\Psi$.  For the sake of examining interactions later,
we also include a scalar field $\Xi$, which $U(1)$ acts on via $\Xi\to
e^{2i\theta}$, to mediate interactions between $\Psi$
quanta\footnote{It is necessary to introduce a field other than
$\Phi$, because the fermion zero mode is chiral, and thus does not
interact with $\Phi$ via the term $g\Phi\Psibar\Psi$.  Additionally,
while the modes of $\Phi$ mix with scalar gravitational degrees of
freedom, the global $U(1)$ symmetry prevents such a mixing of $\Xi$
modes.}, and take the action describing these two fields to be
\begin{equation} \label{eq:toyaction}
\begin{aligned}
\mcs_{\Psi\Xi} = \int\!d^4x\!\int\!dy\sqrt{G}\Big[ i\Psibar\Gamma^M
    \partial_M\Psi - g\Phi\Psibar\Psi &+ \partial^M\Xi^\dagger
    \partial_M\Xi - g'\Phi^2\Xi^\dagger\Xi - u^2 \, \Xi^\dagger\Xi
    - \tau(\Xi^\dagger\Xi)^2 \\
    &- \lambda (\Xi\Psibar\Psi^c + \text{h.c.}) \Big],
\end{aligned}
\end{equation}
where $\Psi^c = \Gamma^2 \Gamma^5 \Psi^*$.  Note that the full
action includes the background given by Eq.~\eqref{eq:gravKink}.

For our background solution to remain stable, $\Xi=0$ must be the
stable solution i.e. Eq.~\eqref{eq:scalar-de} must not have any
negative eigenvalues.  Choosing $u^2>0$ suffices to guarantee this.

\begin{figure}
\centering
\includegraphics[width=0.74\textwidth]{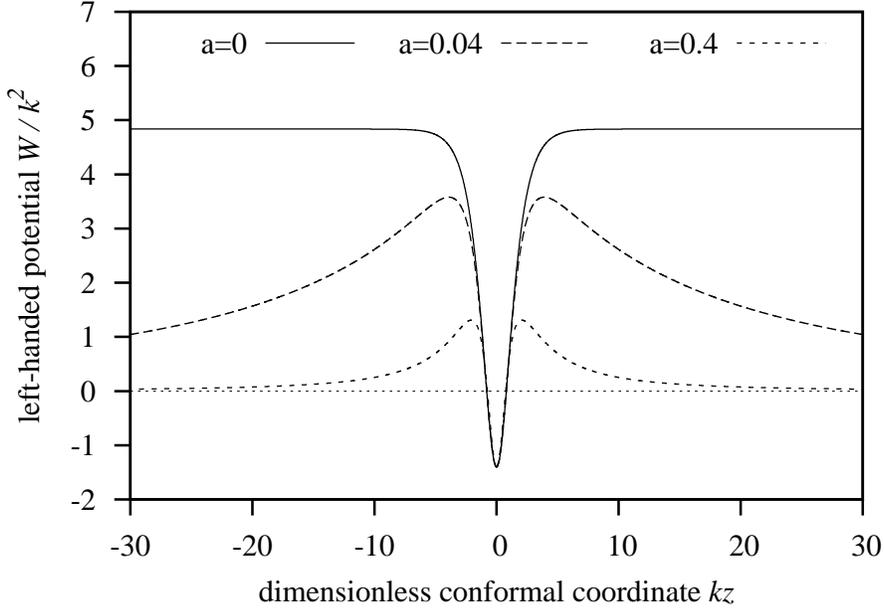}
\caption{
\small
An example of the effective Schr\"odinger potential,
$\tilde{W}_-^{\text{eff}}$,
which traps a left-handed fermion field.  The three graphs
correspond to no gravity ($a=0$), `weak' gravity ($a=0.04$), and
`strong' gravity ($a=0.4$).  The horizontal line is $W=0$.  All
plots have $gD=1.4\, k$.
}
\label{fig:pot}
\end{figure}

The effective Schr\"odinger equation for the fermion field
is found as described in Section~\ref{sec:ferm-grav}.  For
various values of $a$, the resulting effective potential felt by the
left chiral component of the fermion is plotted in Fig.~\ref{fig:pot}.
It does indeed asymptote to zero when gravity is included, implying
the existence of a continuum of arbitrarily light modes.  There is
of course a zero mode which is localised to the brane -- all other
modes are however oscillatory at infinity.

We wish to quantify our argument that the light continuum modes
do not overly influence physics on the brane.  The dominant process
by which they could be detected in our toy model is two zero mode
particles annihilating to produce two continuum particles via exchange
of a $\Xi$ quantum, so this is the process we will consider.  As
explained at the end of Section~\ref{sec:ferm-grav}, a $\Xi$ quantum
produced on the brane will not correspond to a single mass mode, but
will be a wavepacket initially localised on the brane.  The creation
of such a wavepacket on the brane and the ensuing shape of the wavepacket
will be a complicated issue, and not considered here.  Instead we
will simply take $\sech(kz)$ as a typical localised
profile\footnote{Results should be almost identical for any profile
which decays exponentially beyond $|z|\sim 1/k$.} and assume that
a $\Xi$ quantum is produced with the extra-dimensional
wavefunction $\tilde h(z)=\sqrt{k/2}\,\sech(kz)$.
We have computed the Fourier decomposition of $\tilde h(z)$ in terms
of the mass eigenmodes (the eigenfunctions of Eq.~\eqref{eq:scalar-de});
the spectrum is sharply peaked at a mass corresponding to the first
resonant mode, as expected.  We now proceed to calculate the
effective coupling of the fermion modes to this particle in the
dimensionally reduced theory.  This will give us a quantitative
estimate of the likelihood of continuum fermion modes being produced
by on-brane dynamics through s-channel annihilation.  It will also
be a valid estimate for t-channel scattering of localised zero modes
with bulk continuum modes.

\begin{figure}
\centering
\includegraphics[width=0.74\textwidth]{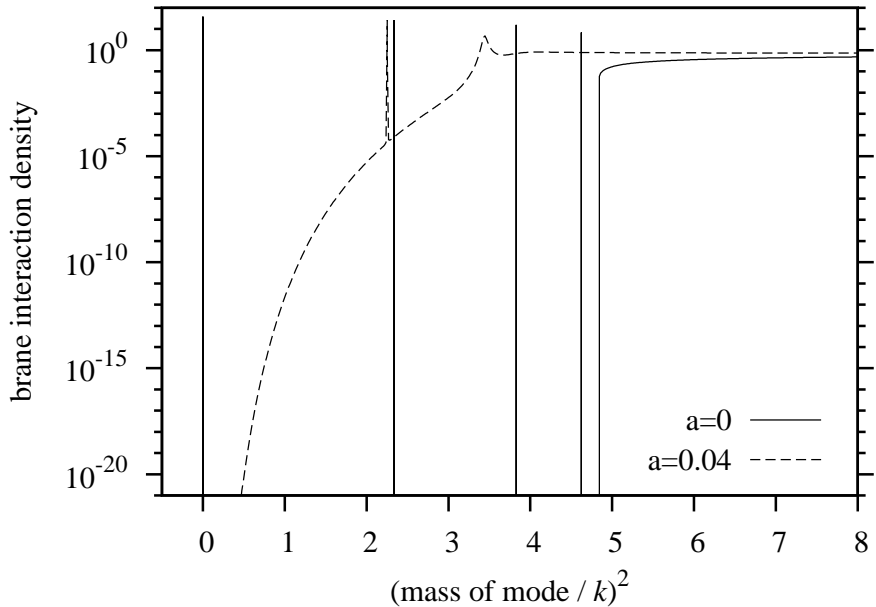}
\caption{
\small
The ``interaction per continuum mode",
$\lambda^{(4)}_n / \lambda \sqrt{k}$, for
continuum modes interacting with a typical bound
mode on the brane.  Both the gravity-free ($a=0$) and
`weak' gravity ($a=0.04$) cases are shown.  The fermion
zero mode remains bound in the presence of gravity (hidden
by the gravity-free plot), whilst a continuum is introduced
for all positive energies.  It is clear that at energies
well below $k$, the continuum modes are essentially
decoupled from those on the brane.  The coupling becomes
relatively strong for energies greater than the maximum of
the effective potential.  The fermion coupling strength
is $gD=1.4\, k$.
}
\label{fig:spec}
\end{figure}

The effective coupling constant between the fermion modes of $\Psi$
and the localised $\Xi$ particle is given by the 5D Yukawa coupling
constant multiplied by the overlap integral of their extra-%
dimensional wavefunctions.  For the fermion mode with extra-%
dimensional dependence $f^n(y)$, the coupling will be
\begin{equation}
\begin{aligned}
\lambda^{(4)}_n &=& \lambda\int\!dy&\, e^{-4\sigma}\, h(y)\,
    \big(f^n(y)\big)^2 \\
&=& \lambda\int\!dz&\,e^{\half\sigma}\tilde h(z) \big(\tilde f^n(z)\big)^2 \\
&=& \lambda\int\!dz&\,e^{\half\sigma}\frac{\big(\tilde f^n(z)\big)^2}
    {\cosh{ky}}.
\end{aligned}
\end{equation}
The results for the case $a=0.04$ are plotted in Fig~\ref{fig:spec},
contrasted with the results in the gravity-free case\footnote{Note that
what is plotted is really ``interaction strength per continuum mode''
with mode energy used on the horizontal axis to label a particular
mode number.  An integral over some finite range of modes is required
to yield a finite on-brane effect.}.  It is clear that the 4D coupling
constants go quickly to zero for modes with masses much less than the
inverse width $k$ of the domain-wall.

\begin{figure}
\centering
\includegraphics[width=0.74\textwidth]{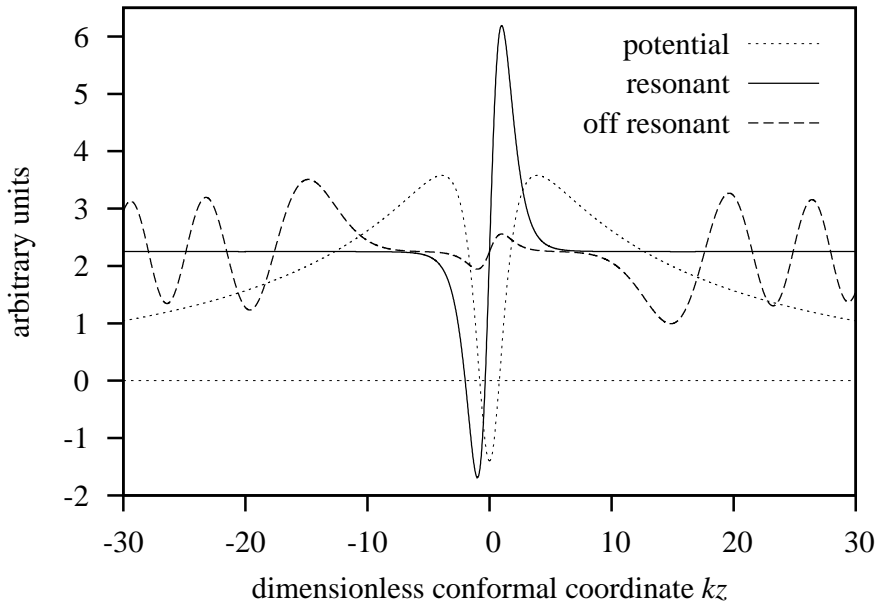}
\caption{
\small
The extra-dimensional profiles of the
resonant mode at $(E/k)^2 \simeq 2.3$, and
a mode off-resonance by $2.0 \times 10^{-4}$
in units of $(E/k)^2$.  We are in the `weak'
gravity case with $a=0.04$. The profiles are not
plotted on the same scale; in reality,
each is normalised to the same amplitude
{\it at infinity} (since the normalisation condition
is dominated by the behaviour of the wavefunction
at infinity).  Thus the contrast is much more
dramatic even than it appears here.
}
\label{fig:res}
\end{figure}

Such behaviour is of course easy to understand, based on the
discussion of Section~\ref{sec:ferm-grav}.  Modes with energy much
less than $k$ see a wide potential barrier preventing them from
penetrating to the brane, where the $\Xi$ particle resides.  At an
energy approximately equal to $k$, we see the first resonant mode,
which does not suffer the generic suppression near the brane.
Continuum modes with energy above the barrier height ($\sim 5k^2$ for
the gravity-free case and $\sim 3.5k^2$ for weak gravity) are
free to roam in the vicinity of the brane, hence their coupling is of
order unity.  We have explicitly plotted the profiles of a resonant
mode and a (slightly) off-resonant mode in Fig.~\ref{fig:res}, to
illustrate the amplification of one and suppression of the other
on the brane.

\section{Conclusion}
\label{sec:concl}

We have shown that when a fermion field is coupled to a gravitating
scalar domain-wall in 5D, the spectrum of the low-energy effective 4D
theory consists of a massless fermion of fixed chirality, and a
continuum of states with all possible masses $m > 0$.  The massless
mode is bound to the brane, while the continuum modes are oscillatory
far from the brane.  A scalar field coupled to the same background
also yields a continuum of arbitrarily light states.

Coupling constants in the low energy theory will be determined by
overlap integrals between the extra-dimensional profiles of the
fields involved.  Generically, continuum modes are strongly
suppressed on the brane, and thus should interact only very weakly
with the zero modes and other localised fields.
We have demonstrated this effect by explicitly computing the overlap
integrals for a typical toy model.
It should be possible within specific models to arrange for the integrated
effects of these modes to be small enough not to contradict known
low-energy phenomenology.  Nevertheless, at higher energies it may be
important to consider the effects of such modes.

There will be a finite number of resonant modes which \emph{will}
manifest on the brane; these are the remnants of the bound states
of the analogue Schr\"odinger system in the non-gravitating case.
The lowest of these modes will have a mass approximately equal to
the inverse width of the domain-wall, which would need to be
sufficiently large in a realistic model.

Note that we have assumed throughout that the 4D metric is Minkowskian.
Cosmologically it may be desirable to allow it to be de Sitter; in this
case by dimensional analysis the continuum matter modes will begin at a mass
$\sim\sqrt\Lambda_4$, where $\Lambda_4$ is the effective cosmological
constant on the brane (this effect is demonstrated explicitly for gravitons
in~\cite{Karch2000}).  For our universe this is sufficiently
small to be negligible for collider phenomenology.

The analysis presented in this paper would form the basis of an
investigation of the low-energy phenomenology of a realistic model;
for example of that presented in~\cite{Davies2007}.


\section{Acknowledgements}

The authors would like to thank Ray Volkas for discussion and support.
DPG was supported by the Puzey Bequest to the University of Melbourne, and
RD by the Henry \& Louisa Williams Bequest to the University of Melbourne.



\end{document}